\documentclass[12pt]{iopart}

\usepackage{graphicx}
\expandafter\let\csname equation*\endcsname\relax
\expandafter\let\csname endequation*\endcsname\relax 
\usepackage{amsmath}
\usepackage{color}
\usepackage[normalem]{ulem}
\definecolor{darkgreen}{rgb}{0,0.6,0}

\newcommand{\benedikt}{B. Y. Mueller}
\newcommand{\picref}[1]{Fig.~\ref{#1}} %Reference for figures
\newcommand{\picrefa}[2]{Fig.~\ref{#1}\,({#2})} %Reference for figures with a,b,c
\newcommand{\eqrefeq}[1]{Eq.~(\ref{#1})} %Reference for figures with a,b,c

\newcommand{\unit}[2]{#1~{\rm #2}}
\begin{document}

\title	{
	   Driving Force of Ultrafast Magnetization Dynamics
	}

\author	{
	  \textrm{\benedikt, T. Roth, M. Cinchetti, M. Aeschlimann, B. Rethfeld}
	}

\address	
	{
	  Department of Physics and Research Center OPTIMAS, 
	  Technical University of Kaiserslautern, Erwin-Schr\"odinger-Str.~46, 67653 Kaiserslautern, Germany
	}

\ead	{
	  bmueller@physik.uni-kl.de
	}

\date{\today}

%----------------------------------ABSTRACT-----------------------------------------------------------------------------------------------------------------
\begin{abstract}
Irradiating a ferromagnetic material with an ultrashort
laser pulse leads to demagnetization on a femtosecond timescale.
We implement Elliott-Yafet type spin-flip scattering, 
mediated by electron-electron and electron-phonon collisions,
into the framework of a
spin-resolved Boltzmann equation.
Considering three mutually coupled
reservoirs, (i) spin-up electrons, (ii) spin-down
electrons and (iii) phonons, 
we trace non-equilibrium electron distributions
during and after laser excitation. 
We identify the driving force for ultrafast magnetization dynamics
as the equilibration of temperatures and chemical potentials
between the electronic subsystems. 
This principle can be used to easily predict the maximum 
quenching of magnetization upon ultrashort laser irradiation in any material, 
as we show for the example of \textrm{3d}-ferromagnetic nickel.
\end{abstract}

\maketitle

%-----------------------NEW INTRODUCTION----------------------------------------------------------------------------------------------------------------------------
Interaction of a femtosecond laser pulse with ferromagnetic metal
is known for 
more than a decade to 
cause ultrafast quenching of the magnetization \cite{Beaurepaire1996}. 
Typical demagnetization
times extracted from experiments on the 3d-ferromagnetic metals Co,
Fe and Ni are between 100\,fs and 300\,fs, depending on the laser
fluence and the specific sample \cite{Koopmans2010,Stamm07,
Cinchetti,Muenzenberg}. From the theoretical side, intense efforts are being
undertaken in order to identify the microscopic mechanisms
responsible for the observed behavior \cite{BigotNMat, Carpene08,
Huebner, Kazantseva2008, Koopmans2005, Krauss, Oppeneer10,
Steiauf2009,Steiauf2010,Steiauf2010_2}. Among the considered mechanisms, Elliott-Yafet
(EY)-type spin-flip scattering \cite{Zutic} is definitely one of the most
implemented 
for the modeling of ultrafast demagnetization. 
The most popular EY-type spin-flip mechanism is electron-phonon
(el-ph) scattering \cite{Koopmans2010, Koopmans2005,Steiauf2009,Steiauf2010,Steiauf2010_2}, while recently
electron-electron (el-el) Coulomb scattering has been considered as
well \cite{Krauss,SteilPRL}. Based on EY-type el-ph scattering, Koopmans et al.\
developed the so-called Microscopic Three-Temperature
Model (M3TM), which has been shown to possess a great predictive
power, explaining 
not only the demagnetization in ferromagnetic
transition metals 
but also the specific demagnetization characteristics of
gadolinium \cite{Koopmans2010}. Electron-electron Coulomb
scattering has been implemented in a dynamical model including
momentum- and spin-dependent carrier scattering \cite{Krauss}, in
the following referred to as the \mbox{el-el} Coulomb (EEC) model. Since
the band structure is taken as input for the calculations, the EEC
model can be successfully applied to 
understand 
the demagnetization
dynamics in materials with peculiar band structure, as recently
shown for the case of half-metallic Heusler alloys \cite{SteilPRL}.
\begin{figure}[b!!!]
\centering
  \includegraphics[
		 width=0.7\linewidth,clip, keepaspectratio,viewport=0.3cm 6.6cm 20.8cm 18.7cm %abschneidebereich aus pdf
		]{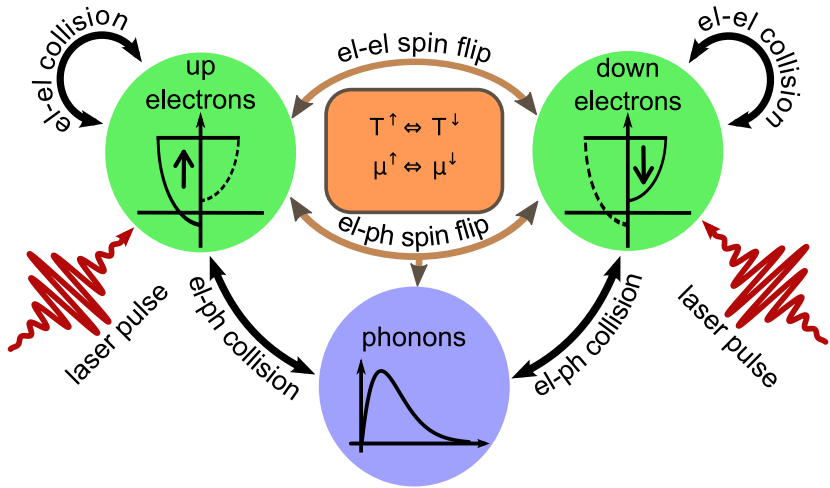}   
  \caption
  {
	Scheme of the spin-resolved Boltzmann equation:
	It separates the phononic and the electronic system,
	the latter subdivided into a majority and a
	minority part. 	
 	Due to the spin-mixing, both electronic systems 
	are coupled by electron-electron and electron-phonon spin-flips.
	In its essence, the driving force for the ultrafast demagnetization
	is determined by an equilibration of temperatures and chemical potentials
	between the two electronic systems. 	
  }
\label{pic:3TM}
\end{figure}

However, a fundamental question remains still unsolved:
What is the ultimate ``driving force''  for ultrafast
demagnetization?

%\rot{To that end}
To answer this question, we introduce  a simplified, yet microscopic kinetic
model, 
describing non-equilibrium electron dynamics in laser-excited 
ferromagnetic metals.
It combines the basic ingredients of the M3TM and the EEC model in the framework of
a \emph{spin-resolved Boltzmann equation}. 
The main idea behind this
approach is schematically depicted in \picref{pic:3TM}. 
%This approach, as schematically depicted in \picref{pic:3TM},
%separates the phononic and the electronic system,
%	the latter subdivided into a majority and a
%	minority part. 	
 	Based on
the EEC model, we describe a ferromagnetic metal as consisting of
three subsystems: the spin-up (majority) electrons,
the spin-down (minority) electrons, and the
phonon system. 
We consider el-el spin-flips and additionally,
in the spirit of the M3TM, we include el-ph mediated spin-flips but
without the need of introducing a separate spin subsystem. 
In order to enable the implementation of complete Boltzmann collision integrals
and to ensure a clear interpretation of the results it is necessary 
%In order to make the implementation possible, it is necessary 
to make simplifying assumptions, like the free electron gas density of states 
%to make the simplifying assumption of a free electron gas density of states 
(DOS) for the two spin subsystems.
The strength of our approach is that it allows to assign a respective 
temperature and chemical potential to up and
down electrons, denoted by $T^{\uparrow,\downarrow}$ and
$\mu^{\uparrow,\downarrow}$. 
We find that the equilibration of these quantities is
the driving force of ultrafast demagnetization.
This equilibration condition can be easily applied to real materials 
with complex DOS to obtain the maximum magnetization 
quenching following laser excitation, as we exemplarily show for nickel. 
%Once we succeeded in our main task, we checked if the model makes 
%correct predictions on real systems and indeed we find a quenching 
%for nickel in the order of the experimental values.
%These quantities, on turn, reveal in a straightforward way the driving 
%force for ultrafast demagnetization.
As a further result, our approach is capable
%Moreover, since our model contains the main ingredients of both, the 
%\rot{M3TM} and the EEC model, it has the potential 
to clarify the relative role played by el-el and el-ph  scattering during ultrafast demagnetization. 

%-----------------------THEORY----------------------------------------------------------------------------------------------------------------------------

In our model, we
%describe the non-equilibrium dynamics of 
%a ferromagnetic metal irradiated with an ultrashort laser pulse.
%To that end, we 
calculate the time- and energy dependent dynamics of the 
considered subsystems on the basis of a kinetic approach
applying complete Boltzmann collision integrals~\cite{Rethfeld}.
The interactions between electrons (el-el), electrons and 
phonons (el-ph) as well
as the absorption of laser energy (absorp)
are described as independent 
collision integrals $\Gamma_{\rm col}$, respectively.
Their sum yields the transient evolutions
of the electronic distribution $f(E)$ and the phononic distribution $g(E)$.
The particular collision integrals 
are generally derived by Fermi's golden rule
\begin{equation}
  \label{eq:goldenrule}
  \Gamma_{\rm{col}}=
      \sum\limits_{\rm all~states}
      \frac{2\pi}\hbar |\langle \varphi_0 | \hat H_{\rm{col}} | \varphi_1 \rangle |^2 \;\delta(E_0-E_1)
  \enspace,
\end{equation}
where $\hat H_{\rm col}$ indicates the Hamiltonian for the considered collision process, 
$\varphi_0$ the initial state with energy $E_0$ and $\varphi_1$ the final state with energy $E_1$.
Such model was developed in Ref.~\cite{Rethfeld} 
to describe non-equilibrium electron and phonon dynamics during and
after ultrafast laser excitation for a non-magnetic material. 
It is capable to account for non-equilibrium 
distribution functions where no real temperatures
are defined. 

First, to apply this approach to ferromagnetic metals 
we include the Stoner model into the existing theory.
This leads to a separated description of the two spin systems (up and down electrons), 
with a constant energy shift by the exchange energy 
$\Delta_{\rm ex}$ between the corresponding distribution functions.

Second,
we include spin-mixing in our approach. 
This allows
a coupling between both electronic reservoirs:
Typically, the orthogonality of the up and down states guarantees that the matrix element 
$\langle \uparrow | \downarrow\rangle$ vanishes and for a spin diagonal 
interaction operator, there will be no possibility
for an electron to change its spin during a collision process.
But due to the spin orbit coupling in solids, 
no pure up or down states are defined, 
but a mixture of both~\cite{Steiauf2009,Steiauf2010}.
In this case, an up-electron, for instance, is described by the state
\begin{equation}
\label{eq:spinmixing}
  |\tilde\uparrow\rangle=a\mid\uparrow\rangle+b\mid\downarrow\rangle \enspace,
\end{equation}
whereby $|b|\ll |a|$. %We assume that $a,b$ do not depend on the electron energy.
As the matrix element of the mixed state
is larger than zero $(|\langle \tilde \uparrow | \tilde\downarrow \rangle|>0)$,
the spin orbit coupling 
allows for spin-flip and thus
induces an interaction between the two spin systems in our model. % up and down electrons.
%The spin-mixing is included 
%together with a spin-independent Hamiltonian for collisions
The collisions including the spin-mixing are thus described by 
\begin{align}
  \begin{split}
  \label{eq:goldenrulespin} 
    \tilde\Gamma_{\rm{col}}
    &=\sum\limits_{\rm all~states}\frac{2\pi}{\hbar} | \langle\sigma_0 | \langle\varphi_{0} |  \hat H_{\rm{col}} | \varphi_{1}\rangle | \sigma_1\rangle |^2\;\delta(E_0-E_1)\\
    &=| \langle\sigma_0|\sigma_1\rangle|^2 \sum\limits_{\rm all~states}\frac{2\pi}{\hbar} \langle\varphi_{0} |  \hat H_{\rm{col}} |\varphi_{1}\rangle  |^2\;\delta(E_0-E_1)\\
    &\equiv|\langle\sigma_0|\sigma_1\rangle|^2 \; \Gamma_{\rm{col}} 
\enspace,
  \end{split}
\end{align}
whereby $\varphi_i$ is the spatial and 
$\sigma_i \in \{\tilde\uparrow,\tilde\downarrow\}$ 
the (mixed) spin wave function of the electron. 

With all the described ingredients, we can now write down 
the \emph{spin-resolved Boltzmann equation}
%which accounts for the exchange energy $\Delta_{\rm ex}$ in the frame of the 
%Stoner model and allows for spin-flip processes 
%by including the effect of spin-mixing.  
%
%
%
\begin{subequations}
  \label{eq:SdBeq}
  \begin{align}
    \label{eq:SdBeq_fermi}
    \begin{split}
    \frac{\partial f^\mu}{\partial t}&=
	\sum\limits_{\sigma,\nu,\lambda}
	|\langle \mu ,\sigma|\nu,\lambda\rangle|^2\;  
	\Gamma_{\rm{el-el}} \\
	&+\sum\limits_{\sigma}|\langle\mu   |\sigma\rangle|^2\;\Gamma_{\rm{el-ph}}
	+\sum\limits_{\sigma}|\langle \mu| \sigma\rangle|^2\;\Gamma_{\rm{absorp}}
    \end{split}\\
    \frac{\partial g}{\partial t}&=
	\sum\limits_{\mu,\sigma}
  	|\langle\mu  |\sigma\rangle|^2\; \Gamma_{\rm{ph-el}}
    \enspace,
    \end{align}
\end{subequations}
with $\mu,\nu,\sigma,\lambda \in \{\tilde \uparrow,\tilde\downarrow \}$ 
indicating the spin states. 
\eqrefeq{eq:SdBeq} consists of three equations for $f^\uparrow$, $f^\downarrow$ and $g$, respectively.
The collision terms $\Gamma_{\rm col}$ 
can be found in Refs.~\cite{Rethfeld,Kaiser}.
Each of the distinct equations is coupled to both others, 
thus the dynamics of 
all three subsystems
%spin-up and spin-down electrons, respectively, 
%as well as of phonons 
mutually depend on each other. 
Since the description allows for spin-flips, the number of particles 
in each electronic reservoir is not conserved. 
Instead, the 
{\em sum} of 
the transient number of particles of spin-up ($n^{\uparrow}$) and spin-down ($n^{\downarrow}$)
is conserved. 
The transient 
magnetization $M$ is calculated through
\begin{equation}
\label{eq:magnetization}	
M(t)=\mu_B(n^\uparrow(t)-n^\downarrow(t))=\mu_B\Delta n(t)
\enspace,
\end{equation}
 whereby $\mu_B$ is the Bohr magneton.
%

%------------------------------RESULTS------------------------------------------------------------------------------
We solve the spin-resolved Boltzmann equation~\eqref{eq:SdBeq}
for a prototype of a ferromagnetic material 
with the Fermi energy
\mbox{$E_F=\unit{8}{eV}$}, 
the effective mass 
$m^*=1.45\;m_e$ 
and the density of states of a free electron gas.
We assume the exchange energy  
$\Delta_{\rm ex}=\unit{2}{eV}$. 
The spin-mixing parameter $b^2 \approx 0.03 $ 
in \eqrefeq{eq:spinmixing}
is chosen according to ab initio calculations for Ni~\cite{Koopmans2010,Steiauf2009,Steiauf2010}.
For the phononic system we apply the Debye model
with $T_D=\unit{724}{K}$ as Debye temperature.
These parameters represent typical values for real metals.
Note that none of them is used as a fit parameter, because it is not in the
scope of this article to reproduce the experimental data, but to
identify the driving force of ultrafast demagnetization.
In this spirit we chose the density of states of a free electron gas 
because essential relations are given analytically and the solution
of the spin-resolved Boltzmann equation can thus be easily interpreted.
%which allows us to explicitly identify the principle 
%mechanisms behind ultrafast magnetization dynamics. 
%
%a 
%in spite of the used
%elementary density of states,
%which, however,
%facilitates the undefiled extraction of 
%
%To clearly distinguish 
%between the effects of the laser and the processes which are 
%independent of photon interactions, we assume a rectangular excitation profile 
%for the first \unit{100}{\femto\second}.
Finally, to clearly investigate the processes after optical excitation, we assume
a rectangular laser profile in time.
%between the effects of the laser and the processes which are 
%independent of photon interactions, we assume a rectangular excitation profile 
%for the first \unit{100}{\femto\second}.

\begin{figure}
\centering
  \includegraphics[
	width=0.7\linewidth,clip, keepaspectratio,viewport=0.65cm 15.8cm 20.15cm 27.9cm %abschneidebereich aus pdf
	]{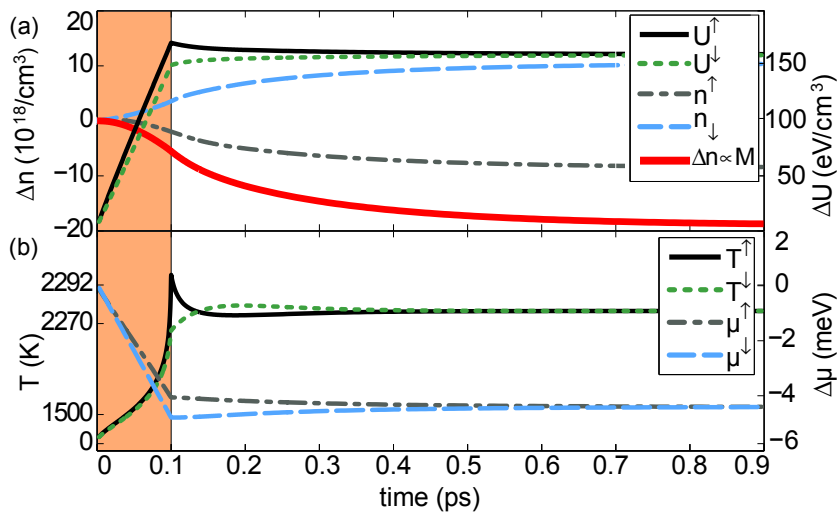}
\caption{
	Simulation neglecting the contribution of phonons. 
	(a) The internal energy $U^{\uparrow,\downarrow}$ and number of 
	particles $n^{\uparrow,\downarrow}$ of each electronic system. 
	(b) Applying \eqrefeq{eqs:implicit} it is possible to define  
	approximated temperatures $T^{\uparrow,\downarrow}$ and chemical potentials $\mu^{\uparrow,\downarrow}$.
	The chemical potentials 
	and the corresponding temperatures 
	approach each other asymptotically.
	}
\label{pic:nophonon}
\end{figure}

\begin{figure}
\centering
  \includegraphics[
		width=0.7\linewidth,clip, keepaspectratio,
		%viewport=1.0cm 15.5cm 19.5cm 29.5cm 
		viewport=2.9cm 16.6cm 15.9cm 26.4cm
		]{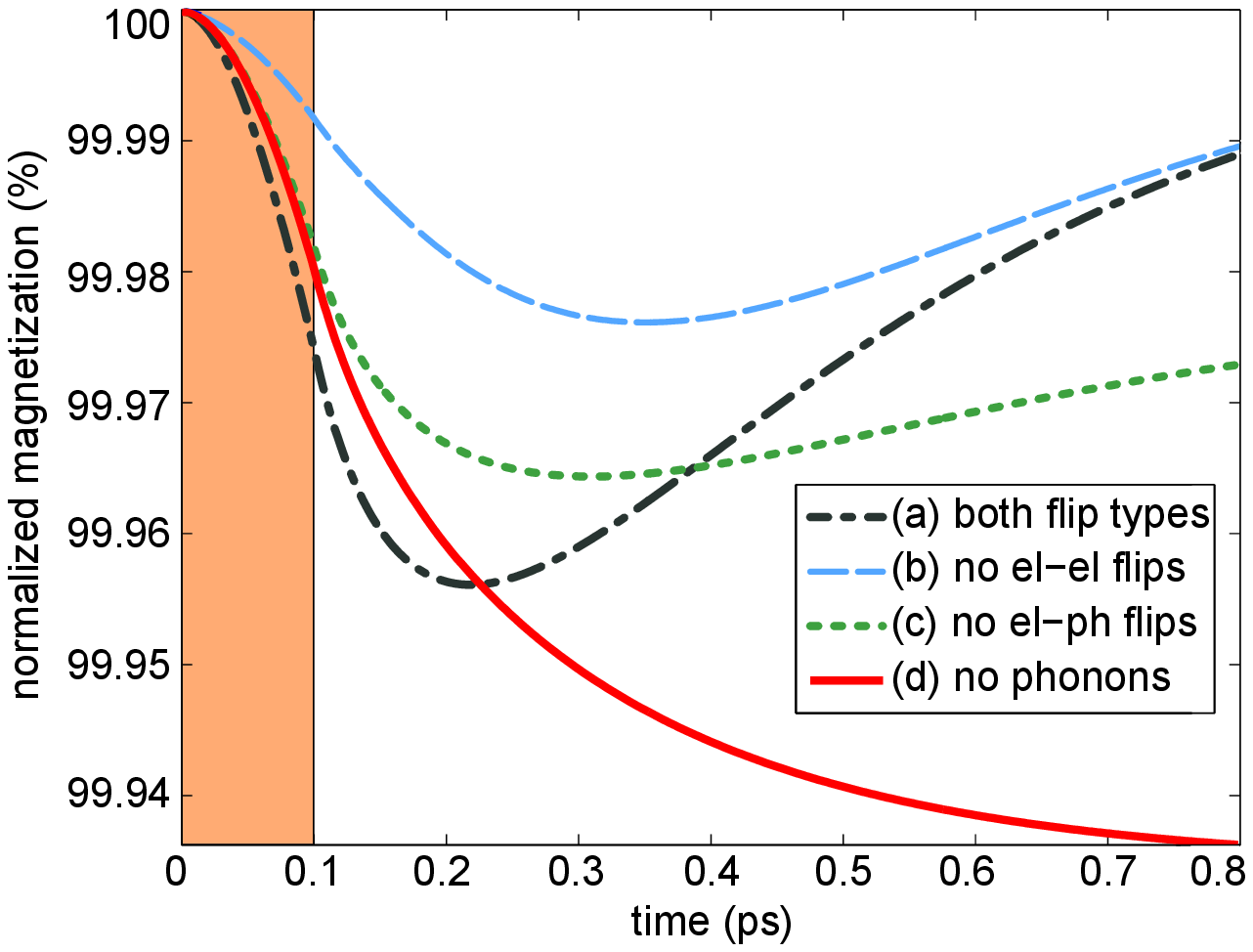}
	\caption{
		%Comparing different types of spin-flip:
		%Phonons are essential for the remagnetization 
		%process (d)
		%and they can increase the effectiveness of the demagnetization (a).
		%If we assume the phonons only as a cooling bath (c) we see
		%a slow recovery of the magnetization.
		%On the other hand, the de- and remagnetization can be achieved 
		%by allowing only electron-phonon- and no electron-electron 
		%spin-flips (b).
		Magnetization dynamics regarding the different scattering processes
		separately. The calculations were performed by applying the 
		spin-resolved Boltzmann equation 
		for a free electron gas.
		}
\label{pic:flips}
\end{figure}

In order to solve the spin-resolved Boltzmann equation~\eqref{eq:SdBeq}
%one has to 
we compute a system of approximately 300 strongly coupled integro-differential equations.
The solution provides the transient non-equilibrium distribution 
function of each electronic reservoir.
It resembles the evolution of the electron distribution in Fig.~1 of
Ref.~\cite{Rethfeld}, but extended by the information on the two spin reservoirs.
In particular, a net spin-flip is observed as a
slight particle exchange between both electron reservoirs 
in the solution of \eqrefeq{eq:SdBeq}
during and after irradiation.
%\begin{figure}[tb]
%\includegraphics[
%	width=\linewidth,clip, keepaspectratio,viewport=2.1cm 11.1cm 19.45cm 24.9cm %abschneidebereich aus pdf
%	]{2_laser}
%	\caption{
%		Solutions of the spin-resolved Boltzmann 
%		equation (see ~\eqrefeq{eq:SdBeq}) by using 
%		a \unit{100}{\femto\second} rectangular shaped laser pulse.
%		Main: Transient magnetization by varying the laser fluence
%		from \unit{0.13}{\milli\joule\per\centi\meter^2} 
%		to \unit{0.64}{\milli\joule\per\centi\meter^2}.
 %		Inset: The maximum quenching $q$ increases linearly with the fluence $F$ of the exciting laser pulse.}
%\label{pic:fluence}
%\end{figure}
%----mu-T---
Figures \ref{pic:nophonon} and \ref{pic:flips} were
calculated for a \unit{100}{fs} laser pulse 
with wavelength $\lambda = \unit{630}{nm}$
and a total absorbed fluence of $F = \unit{0.13}{mJ/cm^2}$.
In Fig.~\ref{pic:flips}, which will be discussed later
in detail, curve (a)
shows the transient magnetization (normalized to the initial 
magnetization) obtained with the spin-resolved Boltzmann equation 
~\eqref{eq:SdBeq} and applying Eq.~\eqref{eq:magnetization}.
%
%Figure \rot{xxx.c)}%\ref{pic:fluence} 
%shows the transient magnetization (normalized to the initial %magnetization $M(t=0)$)
%obtained through \eqrefeq{eq:magnetization}. 
%%for different laser fluencies.
We obtain a demagnetization time of
\mbox{$\tau_M\approx \unit{100}{fs}$} after laser irradiation.
Remagnetization occurs on  a picosecond timescale.

%and a linear increase of the maximum quenching $q$ with the fluence $F$
%(inset of \picref{pic:fluence}).
%This result is in qualitative accordance with experiments \cite{FluenceLinear}.
%Such linear dependence has been observed experimentally \cite{FluenceLinear},
%but with higher absolute values of the quenching. 
%The reason for the small quenching obtianed in our model 
%lies in the simplified density of states as will be explained below.

%---------------Magnetismus------------------

Generally, when a system is excited out of its equilibrium state, it tends
back to a new equilibrium. An example is the 
relaxation between electrons and phonons after ultrafast laser excitation. 
Here, the initially heated electrons transfer energy to the lattice 
in a way that both end at a new, higher temperature. 
The nonequilibrium of temperatures between both systems drives
this energy exchange.
However, what is the driving force of particle exchange
between spin-up and spin-down electrons, which,
according to Eq.~\eqref{eq:magnetization}, 
is the
basis for magnetization dynamics? 
%The corresponding equilibrium condition for particle exchange
%is an equilibrium of chemical potentials.
For such a system, allowing for energy {\em and} particle exchange,
the equilibrium condition refers to temperature {\em and} chemical 
potential and reads
%
%\rot{Both together, we obtain the equilibrium condition 
%for a magnetic system as}
\begin{equation}
  \label{GLS}
	T^ \uparrow =T^\downarrow \quad \mbox{  and }\quad
	\mu^ \uparrow=\mu^\downarrow+ \Delta_{\rm  ex}
        \enspace.
\end{equation}
Exploiting the simplifications possible for a free electron gas, the chemical potentials 
can be explicitly expressed by the Sommerfeld expansion:
\begin{align}\label{taylor}
	\mu^{\uparrow,\downarrow}(T^{\uparrow,\downarrow}) 
=  E_F(n^{\uparrow,\downarrow})
\left[1-\frac{\pi^2}{12}\left(
\frac{k_BT^{\uparrow,\downarrow}}{E_F(n^{\uparrow,\downarrow})}
\right)^2+\dots \right]
\enspace.
\end{align}
After optical excitation, the temperatures $T^\uparrow$ and 
$T^\downarrow$ will be increased. 
Let us assume that $T^\uparrow$ and $T^\downarrow$ were already equalized.
Then Eq.~\eqref{taylor} clarifies that
in order to equal also the chemical potentials 
$\mu^\uparrow$ and 
$\mu^\downarrow$, 
%as required in the equilibrium conditions
%\eqref{GLS}, 
a change of $E_F(n^{\uparrow,\downarrow})$, mediated by 
a change of $n^\uparrow$ and 
$n^\downarrow$, is required.
Thus, it is the equilibration of chemical potentials of the two electronic
subsystems, which leads to a particle exchange between both 
reservoirs and thus provides the 
ultimate
driving force for ultrafast magnetization
dynamics.
%Thus we see that both conditions in \eqref{GLS}
%can be fulfilled only, if a change of $n^\uparrow$ and 
%$n^\downarrow$ leads to a change of $E_F(n^{\uparrow,\downarrow})$.

%\gruen{\em jetzt muessen wir aber doch ein bisschen aufpassen. 
%Oben haben wir gesagt, weil wir particle exchange haben, muessen
%wir auf das chemische Potential gucken. 
%Jetzt sagen wir, weil wir auf das chemische Potential gucken, 
%muessen wir particle exchange haben. 
%Ist das schluessig? Wie formulieren wir das, ohne dass die Katze
%sich in den Schwanz beisst? 
%Wodurch ist nun wirklich festgelegt, dass unsere chemischen Potentiale
%gleich sein sollen? Ist das eben einfach so? Ist das immer so??}
%
%------------------------------------No Phonons---------------------------------------------------------------------------------------------------------------

\label{sec:nophonons}
Our model allows to verify the conclusion of the driving force behind the demagnetization process 
because we can selectively suppress different scattering channels in \eqrefeq{eq:SdBeq}.
%contributions of el-el and el-ph scattering
%for the demagnetization process
%we suppress on purpose different scattering channels in .
%
%In the following, we focus on the interaction of the two electronic reservoirs 
In the following, we discard the phononic influence 
%in \eqrefeq{eq:SdBeq} 
by setting 
$\Gamma_{\rm el-ph}=\Gamma_{\rm ph-el}=0$.
The transient spin-resolved density $n^{\uparrow,\downarrow}$ and internal energy $U^{\uparrow,\downarrow}$
are depicted in dependence on time
in \picrefa{pic:nophonon}{a} together with the difference
$\Delta n$, being proportional to the magnetization.
Due to the higher electron density of majority electrons
the amount of energy absorbed by the spin-up electrons is 
larger than for the spin-down electrons.
For each moment in time and both spin directions
we may define a temperature $T^{\uparrow,\downarrow}$ and 
a chemical potential $\mu^{\uparrow,\downarrow}$
of the current distribution function.
Both quantities are found through the
implicit integral equations 
%built by the zeroth and second moments 
of the corresponding Fermi distribution $f_F$: 
\begin{subequations}
  \label{eqs:implicit}
  \begin{align}
        n^{\uparrow,\downarrow}(t)&\stackrel{!}{=}\int
		f_F^{\uparrow,\downarrow}
		  (E,T^{\uparrow,\downarrow},\mu^{\uparrow,\downarrow}) 
			\, \mbox{DOS}(E)  \;\textrm{d} E\enspace,\\
	U^{\uparrow,\downarrow}(t)&\stackrel{!}{=}\int
		f_F^{\uparrow,\downarrow}
		  (E,T^{\uparrow,\downarrow},\mu^{\uparrow,\downarrow}) 
			\, \mbox{DOS}(E) \, E \;\textrm{d} E
        \enspace.
  \end{align}
\end{subequations}
The resulting temperatures $T^{\uparrow,\downarrow}$ and chemical potentials $\mu^{\uparrow,\downarrow}$ are 
depicted in \picrefa{pic:nophonon}{b}.
Due to laser excitation, both temperatures
increase, while the chemical potentials drop. 
When the laser is turned off (at $t=\unit{100}{fs}$)
the two spin systems differ in
temperatures {\em and} chemical potentials.
Since this difference is small for the assumed case of 
free electron like density of states (second order term in \eqrefeq{taylor}), the quenching of magnetization 
is also considerably lower than observed in real materials.
Subsequently, the reservoirs equilibrate until
a same temperature $T$ {\em and} chemical potential $\mu$ are reached. 
According to Eq.~\eqref{GLS}
{\em both} constraints have to be satisfied to %quantities has to equilibrate to 
reach an equilibrium between the electronic systems.
\picrefa{pic:nophonon}{b} verifies that the temperatures
and chemical potentials equilibrate simultaneously.
%Note that both, 
%$T^{\uparrow,\downarrow}$ and $\mu^{\uparrow,\downarrow}$,
%equilibrate simultaneously.}
Comparing with 
\picrefa{pic:nophonon}{a}, 
we find that also the transient magnetization, 
being proportional to $\Delta n$, 
reaches its asymptotical value 
on the same timescale.
%according to \eqrefeq{eq:magnetization},
%reaches asymptotically a finite 
%value $M_{\rm asymp}^{\rm el}$ smaller than that of the unperturbed 
%state. 
%The decrease occurs on the same timescale
%where $T^{\uparrow,\downarrow}$ and $\mu^{\uparrow,\downarrow}$ 
%equilibrate.
%
%
% the transient magnetization 
%according to \eqrefeq{eq:magnetization},
%reaches asymptotically a finite 
%value $M_{\rm asymp}^{\rm el}$ smaller than that of the unperturbed state. 
%
%Hence, the non-equilibrium of temperatures and chemical potentials for 
%excited electrons forces
%not only an energy exchange but also
%a particle exchange between the spin reservoirs. 
%
%
In its essence, the particle exchange affected by the difference
in chemical potentials provides the driving force
for ultrafast magnetization.

%\begin{figure}[tb]
%  \includegraphics[
%	width=0.6\linewidth,clip, keepaspectratio,viewport=0.65cm 14.0cm 20.15cm 27.9cm %abschneidebereich aus pdf
%	]{../pics/stoner_inkscape}
%\caption{
%	Einfach mal ein Test. Aber das Bild nimmt zu viel Platz weg.
%	}
%\label{pic:stoner}
%\end{figure}

%The value of $M_{\rm asymp}^{\rm el}$
%which thus depends only on an equilibrium electron temperature \mbox{$T_e\equiv T^\uparrow=T^\downarrow$}
%can be obtained for any density of states without 
%solving the spin-resolved Boltzmann equation \rot{but using the idea \blau{\em suche besseres Wort} of the equilibration 
%of $T$ and $\mu$}.
%In case of starting at room temperature and 
%ending up at an electron temperature of $T_e = 3000~\rm K$, 
%we predict a maximum quenching 
%for nickel of about 30\% using its real band structure \cite{Zhigilei}.
In order to estimate the effect for more realistic systems, 
we choose nickel and solve the equilibrium condition
\eqref{GLS} applying the implicit equations \eqref{eqs:implicit}
with the real density of states for Ni \cite{Zhigilei}.
Note that this can be done without explicitly solving the 
spin-resolved Boltzmann equation.
%For the example of nickel with its real density of states from 
%Ref.~\cite{Zhigilei}
In case of starting at room temperature and 
ending up at an electron temperature of $T_e\equiv T^\uparrow=T^\downarrow = 3000~\rm K$
a maximum quenching of $30\%$ is found.
Such transient electron temperatures are typical
for ultrafast laser excitation
of metals below melting threshold \cite{Rethfeld,Ivanov}.
Thus, applying the main result of our simplified model to realistic systems leads to 
quantitative agreement with experiments,
demonstrating the solidity of our findings.

%\rot{\sout{Finally}}Note that 
%the maximum possible quenching is determined solely 
%by the energy increase in the electronic reservoirs which 
%depends on the energy density of the laser pulse. 

%-----------------------------------------Different Spinflip types----------------------------------------------------------------------------------------------------------
To show the great potential of the spin-resolved Boltzmann equation, 
we now exemplarily compare the relative effects of
el-el spin-flips and el-ph spin-flips
on the demagnetization process. We again restrict ourselves to the 
case of a free electron gas.
Figure \ref{pic:flips}
represents the calculated magnetization dynamics obtained with 
the spin-resolved Boltzmann equation \eqref{eq:SdBeq},
when different collision processes are regarded separately.
The red solid line (d) corresponds to the red solid line 
in \picrefa{pic:nophonon}{a}, determining the magnetization dynamics
when the influence of the phonon bath is completely discarded. 
Including the cooling by phonons but disregarding el-ph spin-flips 
(green dotted line (c)),
we see a slight recovery of magnetization.

Thus, phonon cooling is essential for the 
\textit{re}mag\-ne\-ti\-zation process.
%, due to their cooling effect.
However, phonons also strengthen the magnetization dynamics:
If we additionally allow for phonon mediated spin-flips we find the 
black dash-dotted curve (a). 
%of magnetization dynamics. 
The demagnetization occurs faster than in the case where only 
electron collisions mediate spin-flips, while the remagnetization 
is strongly accelerated by phonon mediated spin-flips. 
Hence, 
phonons can act like a catalyst for the ultrafast magnetization
process. 
The blue dashed line (b) is calculated excluding electron-electron-collision mediated 
spin-flips. In this case, the quenching of magnetization will be less. 
Comparing the simulations shown in \picref{pic:flips}
%with both flip types (a) with those
%including only 
%electron mediation (d) or phonon mediation (b), respectively,
we conclude that for our system electron mediated spin-flips are important 
mainly during the {\em de}magnetization phase, while 
phonon mediated spin-flips dominate the {\em re}magnetization phase
and increase the effectiveness of the demagnetization.

%----------------------------------------CONCLUSION-----------------------------------------------------------------------------------------------------------
In conclusion, we described the ultrafast demagnetization 
with the spin-resolved Boltzmann equation.
%The solutions show the well known ultrafast demagnetization behavior.
%Like in experiments, the minimum of the magnetization is reached on
%a femtosecond time scale and the maximum quenching depends linearly on the laser fluence.
We identified the driving force of the demagnetization process 
%is given by
in the equilibration of the temperatures {\em and} chemical potentials 
of the up and down electrons.  
As shown for the example of Ni this equilibrium condition provides 
a possibility to easily estimate the maximum quenching for any 
ferromagnetic material.
Our approach is also capable to investigate the role of different scattering processes
during ultrafast demagnetization dynamics.
%We revealed that the deeper reason of the demagnetization 
%is the non-equilibrium of both electronic systems whereas the scattering channels,
%wether it is el-ph or el-el mediated spin-flips, determine only the timescales.

%The separation of the different interaction types
%has revealed that phonons can significantly amplify and accelerate the effect
%of de- and remagnetization, while solely the 
%energy content of the
%electrons, or rather $T_e$,
%define the 
%magnetization in electronic equilibrium.
%The latter fact was utilized to predict 
%the maximum possible quenching. 

%---------------------------------------------ENDE------------------------------------------------------------------------------------------------------

\section*{Acknowledgement}
%-------------------------------------------------------------------------------------------------------------------------------------------------- 

We acknowledge fruitful discussions with H. C. Schneider.
Financial support of the Network Project "UltraMagnetron" of the European Union 
and the Deutsche Forschungsgemeinschaft through the 
GRK 792 "Nonlinear Optics and Ultrafast Dynamics"
and the Emmy Noether project RE 1141/11-1 "Ultrafast 
Dynamics of laser-excited Solids" is gratefully acknowledged.

\section*{References}
%--------------------------------------------------------------------------------------------------------------------------------

\end{document}